\documentclass[aps,prl,twocolumn,justify,superscriptaddress,groupedaddress,nofootinbib,preprintnumbers]{revtex4}
\usepackage{graphicx}  
\usepackage{bm}        
\usepackage{amssymb}   
\usepackage{subfig}
\usepackage{enumerate}

\newcommand{\nc}{\newcommand}
\nc{\hf}{\frac{1}{2}}
\nc{\bea}{\begin{eqnarray}}
\nc{\eea}{\end{eqnarray}}
\nc{\be}[1]{\begin{equation} \mbox{$\label{#1}$}}
\nc{\ee}{\vspace{0.1cm}\end{equation}}
\nc{\eq}[1]{\mbox{Eq.\ (\ref{#1})}}
\nc{\fig}[1]{\mbox{Fig.\ (\ref{#1})}}

\def\GeV{{\rm \ GeV}}

\begin{document}

\title{Curvaton model completed}
\author{Kari Enqvist}
\email{kari.enqvist@helsinki.fi}
\author{Rose N. Lerner}
\email{rose.lerner@helsinki.fi}
\author{Olli Taanila}
\email{olli.taanila@helsinki.fi}
\affiliation{Physics Department, University of Helsinki and Helsinki Institute of Physics\\
P.O. Box 64, FI-00014, Helsinki, Finland}

\preprint{HIP-2011-13/TH}

\begin{abstract}

In an inflationary cosmology, the observed primordial density perturbation could come from the quantum fluctuations of another light ``curvaton'' field, rather than the inflaton. In this case, it is essential that the curvaton decays, converting its perturbation to an adiabatic perturbation. For the first time, we consistently account for this decay in the simplest curvaton model $V(\sigma) = \hf m^2\sigma^2$ and point out that it gives rise to an important logarithmic correction to the potential. Moreover, the potential will also receive a correction from the thermal bath. As a consequence, the dynamics of the curvaton are substantially altered compared to the original model in the majority of the parameter space. It will therefore be necessary to re-calculate all the predictions of the original model.

\end{abstract}

\maketitle

We observe in the cosmic microwave background (and infer from large scale structure) an adiabatic, superhorizon, almost scale-independent spectrum of primordial perturbations. A common hypothesis is that these perturbations originated as quantum fluctuations of a light scalar field when the whole Universe was inside the horizon during inflation. This field could be the inflaton itself, or another light field, called the curvaton $\sigma$ \cite{Enqvist:2001zp,Lyth:2001nq,Moroi:2001ct}, whose energy density $\rho_\sigma$ is subdominant during inflation. With the forthcoming Planck Surveyor Mission precision data it will be possible to test not only different models of inflation but also alternative models for the origin of the primordial perturbation.

In the curvaton scenario inflation is driven by some field whose contribution to the curvature perturbation is negligible. There is also another field, the curvaton, which is very subdominant during inflation, with some initial field value $\sigma_\ast$. During inflation the curvaton acquires perturbations from its quantum fluctuations. After inflation ends and the inflaton decays, the universe is filled with radiation. However, the curvaton perturbation continues evolving, with the curvaton field oscillating in its potential until it too decays into radiation. At this time the curvaton perturbation is converted into an adiabatic curvature perturbation of the Universe. The amplitude of the final perturbation, which should match observations, depends on both how long the curvaton oscillates before it decays, and on the shape of the potential. At the time of the decay, the curvaton may have become the dominant energy form or it may still be subdominant.

The simplest curvaton model is given by the quadratic potential
\be{sosimple}
V_0=\frac 12 m^2\sigma^2~,
\ee
where $m$ is the curvaton mass. Such models have been extensively studied (see for example \cite{Dimopoulos:2003ss} and references within). Models
with renormalisable and non-renormalisable self-interacting
terms in the Lagrangian have also been considered (see \cite{Enqvist:2010dt} and references within for details). However, since the curvaton must decay, the curvaton must couple to other fields, and the couplings to other degrees of freedom will generate corrections to the potential, rendering the naive treatment of a purely quadratic potential inconsistent in large areas of the parameter space.

In this letter, we take the simplest realisation \eq{sosimple} of the
curvaton model and consistently implement the requirement of curvaton
decay. We consider the perturbative decay of the
curvaton, although we note  that the decay could instead occur via
parametric decay \cite{Enqvist:2008be,BasteroGil:2003tj}.
The curvaton, a scalar field, can be coupled in principle to either another scalar field $\phi$ or a fermion field $\psi$. These interactions should be present in the bare Lagrangian of the curvaton, proportional to
\[ g^2\sigma^2\phi^2 \quad \mathrm{or} \quad g\sigma \bar{\psi}\psi \; ,\]
with some coupling constants.
The perturbative two body decay width to massless particles is then given
by
\be{decay1} \Gamma = \gamma\frac{g^2 m}{64\pi^2}~,
\ee
where $g$ is the coupling strength and $\gamma$ measures the number of available decay channels,
higher order effects and can account for many body decays (thus depends on the particle content). Unless there are a large number of degrees of freedom present, $\gamma$ can be expected to be of $\mathcal{O}(1)$, and thus we take $\gamma=1$ here for simplicity.

The curvaton model now has two parameters, the mass $m$ and the coupling constant $g$ (or $\Gamma$), and two initial conditions, $H_*$ and $\sigma_*$. One of these parameters should be determined from the requirement that the curvaton produces a curvature perturbation with the observed amplitude, $\zeta \simeq 10^{-5}$. Furthermore as we will discuss below, the initial field value $\sigma_\ast$ can be constrained by probabilistic arguments.

The existence of couplings to other fields implies quantum
corrections of the well-known Coleman-Weinberg type for the effective potential of the curvaton{\footnote{In principle, there will be gravitationally induced corrections to the potential. These will either be proportional to $R$ (which is zero in a radiation dominated background, as considered here) or higher order operators suppressed by $M_p$. Therefore these corrections do not affects the dynamics after inflation.}}.
The effective one-loop potential for the model (with effective coupling $g$ to
effectively massless fields) is given by
\be{deltaV}
\Delta V(\sigma) = \beta \frac{g^2\sigma^4}{64\pi^2}
\left[\log\left(\frac{g\sigma^2}{\mu^2}\right) - \frac{3}{2}\right]
\ee
where $\beta$ is some number depending on the number of fields $\sigma$ is
coupled to, and $\mu$ is the renormalization point. A coupling to bosons gives a positive contribution to $\beta$, while a coupling to fermions gives a negative contribution. For now, we take $\beta=1$ for simplicity.

After inflation ends, the background is dominated by radiation. Because of the thermal background there is also a thermal correction to the potential. To lowest order (for $T\gg m$) it is given by
\be{thermalcorr}
V_T=\alpha g^2T^2\sigma^2~,
\ee
where $\alpha$ again depends on the number of fields the curvaton couples to; here we take $\alpha=1$.
Right after inflation (assuming instant thermalization of the inflaton decay products) the temperature is given by $T\approx({M_pH_*})^{1/2}$.
Because of the thermal bath, the curvaton could also decay by absorption; however, there are kinematical constraints that prohibit
absorption if the curvaton couples to only one field (or, in case of fermions, if the thermal masses of left handed and
right handed fermions are equal) \cite{absorption}. Here we neglect absorption for simplicity.

\begin{figure}[!tb]
\includegraphics[width=\columnwidth, angle=0]{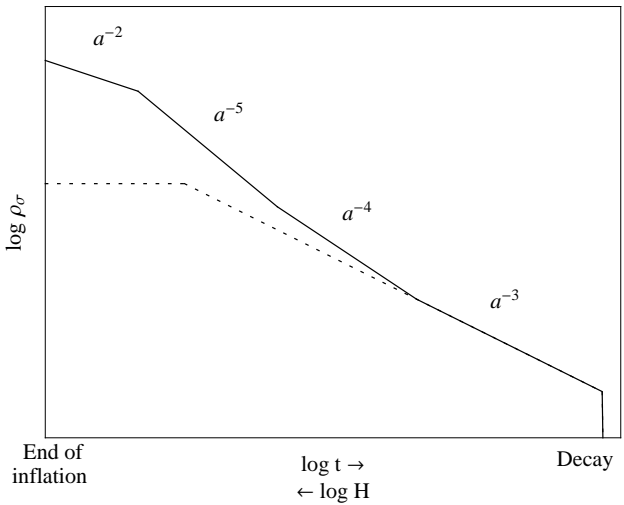}
\caption{\footnotesize{A schematic demonstration of the evolution of the curvaton energy density for the full potential \eq{Veff} (solid) and for the original model \eq{sosimple} (dashed). In order to obtain the same value of $\zeta$, the full potential requires a much larger initial energy density than is obtained by considering only the original model. This difference can be several orders of magnitude. In the case of the full potential, the evolution begins as slow roll in $V_T$ ($\propto a^{-2}$), followed by oscillations in $V_T$ ($\propto a^{-5}$), oscillations in $\Delta V$ ($\propto a^{-4}$) and finally oscillations in $V_0$ ($\propto a^{-3}$).}
\label{law}}
\end{figure}

The total one-loop effective potential is thus
\be{Veff}
V_{eff}(\sigma)=\frac 12 m^2\sigma^2+\Delta V(\sigma)+V_T(\sigma,T)~.
\ee
Hence the curvaton does not oscillate in a harmonic potential. This modifies the effective equation of state and the scaling law of the energy density of the curvaton, as is demonstrated in \fig{law}. The figure shows the evolution of $\rho_\sigma$ where the solid line is for the case where the curvaton energy density is dominated in turn by the thermal correction $V_T$ (in slow roll), $V_T$ (oscillating), $\Delta V$ (oscillating) and finally oscillations in $V_0$ and an instantaneous decay. In comparison, the evolution ignoring both decay and thermal corrections is shown with a dashed line. In order to match the observed curvature perturbation $\zeta$, the curvaton energy density $\rho_\sigma$ must have a certain value at the time of decay. Thus it is clear that the two scenarios produce the same curvature perturbation $\zeta$ only with very different initial conditions. As a consequence the ramifications of the simplest curvaton model are not revealed by the study of the quadratic form \eq{sosimple}, as we will now discuss.

\begin{figure*}[!htb]
    \subfloat[\footnotesize{This figure is plotted with $H_\ast = 10^{10}\GeV$. The WIMP limit (condition (v) in the text) requires $\Gamma$ to be above the dashed line.}]
    {\label{H10} \includegraphics[width=0.95\columnwidth, angle=0]{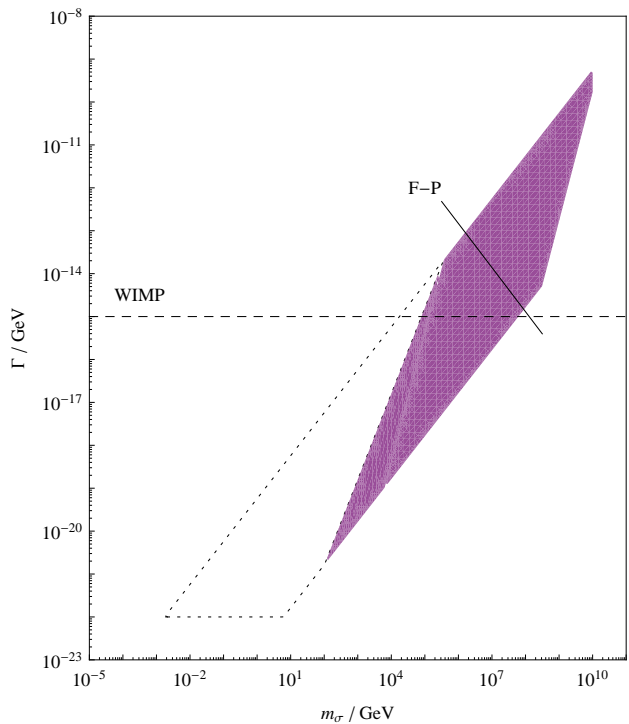}}
    \hspace{0.3in}
    \subfloat[\footnotesize{This figure is plotted with $H_\ast = 10^{12}\GeV$. Note the different scale of the axes compared to (a).}]
    {\label{H12} \includegraphics[width=0.978\columnwidth, angle=0]{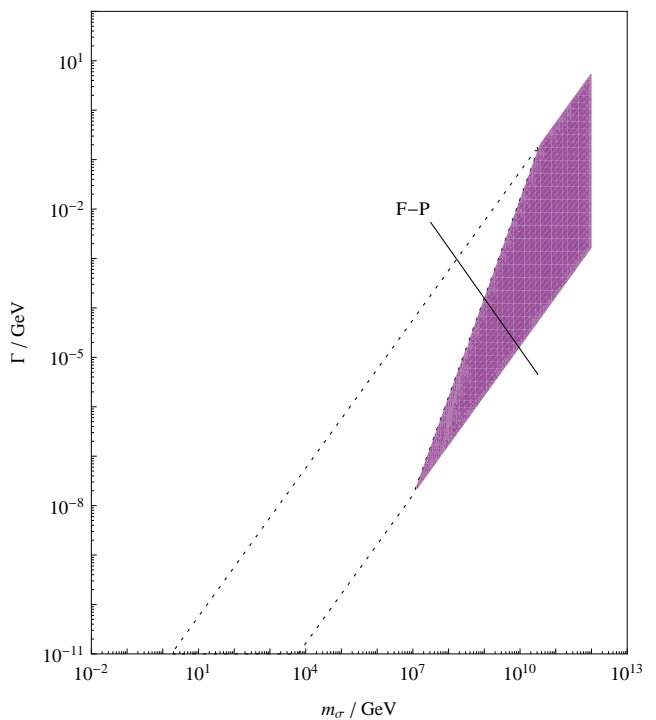}}
    \caption{\footnotesize{The coloured region shows the allowed parameter space. If the contributions of $\Delta V$ and $V_T$ were neglected, the allowed region would extend to include the area bordered by the dotted line. In (b) this extends beyond the scope of the figure. Below the solid line (marked ``F-P'') is the region with high probability (assuming a long period of inflation).} \label{pspace}}
\end{figure*}

In order to further illustrate the importance of using the full effective potential \eq{Veff}, we consider the parameter space in terms of the observables $m$ and
$\Gamma$. For each combination $\{m,\Gamma,H_\ast\}$,  we choose the initial
field value $\sigma_\ast$ such that the curvature perturbation matches the observed value $\zeta\simeq 10^{-5}$ where possible. It is given by
\be{zeta} \zeta =
\frac{H_{\ast}r_{dec}}{3\pi\sigma_\ast} \; ,
\ee
where
\be{rdecaydef}
r_{dec} \equiv \left.\frac{3\rho_\sigma}{3\rho_\sigma +
4\rho_\gamma}\right|_{decay}\;.
\ee

In our analytical estimates, we ignore the evolution of the factor $\left(\log\left({g\sigma^2}/{\mu^2}\right)
- {3}/{2}\right)$ in \eq{deltaV} since it changes only logarithmically, and replace with unity. (We have checked that its effect on the
scaling law of the radiation is negligible.) We also assume sudden decay
when $H=\Gamma$ and that only one of the terms in the potential dominates at a time.
The curvaton is assumed to be frozen $\sigma = \sigma_\ast$ until $H^2 = V''$, where
`$\ast$' denotes the end of inflation throughout this letter.
We further approximate $r_{dec}$ as ${\rho_\sigma}/{4M_p^2 H^2}$.

To ensure a consistent scenario, we apply various constraints to the parameter space:
\begin{enumerate}[(i)]
\item the initial field value should be sub-Plankian ($\sigma_\ast < M_p$),
\item the curvaton should be light during inflation $( m_{eff} < H_{\ast})$,
\item the curvaton energy density should be initially negligible ($r_\ast \ll 1$),
\item the curvaton should decay before nucleosynthesis ($\Gamma \gtrsim 10^{-22}\GeV$)
\item the curvaton should decay before dark matter freezeout. If dark matter is a WIMP which decouples at $T_{DM} \sim {\cal O}(10)\GeV$, then $ \Gamma > 10^{-15} \GeV$. This bound depends on the model of dark matter and could be relaxed to allow smaller $\Gamma$.
\item non-Gaussianity should be below the current observational bound, usually parameterised by $f_{NL} \simeq {5}/({4r_{dec}}) \lesssim 74$ \cite{Komatsu:2010fb}.
\item the decay width should be at least as large as the gravitational decay width ($\Gamma \gtrsim \lambda m^3 /M_p^2$ where we conservatively take $\lambda = 10^{-3}$)
\item the model should generate a curvature perturbation which matches observations ($\zeta \sim 10^{-5}$)
\end{enumerate}
Constraints (i) and (iii) do not impact the parameter space as the other constraints are stronger. 

If inflation lasts long enough, we can also constrain the initial condition for the curvaton from the Fokker-Planck limit. It can be shown that in the
limit of an infinite duration of inflation, the initial field value
will be limited by the equilibrium probability distribution
\cite{Starobinsky:1994bd,Enqvist:2009zf}:
\be{prob1} P = \frac{1}{N}
\exp\left(-\frac{8\pi^2}{3H_\ast^4}V(\sigma_\ast)\right) \ee
which is the equilibrium solution of a Fokker-Planck equation balancing
the classical dynamics and quantum noise. This is valid under the assumption that the Hubble parameter evolves slowly in the last 60 or so e-foldings of inflation. We consider the region $P\geq 10^{-3}/N$
to have high probability under the above assumption. To be precise, for a finite number of e-folds, one should also consider the rate of relaxation into the asymptotic distribution \eq{prob1}; we do not address this issue here{\footnote{The potential in \eq{prob1} does in principle contain gravitationally induced quantum corrections. For simplicity, we do not consider these here and the Fokker Planck limit should be taken as indicative.}}.

In \fig{pspace}, we show the available parameter space for two representative values of $H_\ast$: $H_\ast = 10^{10}\GeV$ and $H_\ast = 10^{12}\GeV$. The results are qualitatively the same even if values of $\alpha$ and $\beta$ as large as 100 are used (the figures show $\alpha =\beta = 1$). The coloured regions show the allowed parameter space for the model. This is bounded from above by the requirement of generating the observed $\zeta$ (viii) and from below by the observational bound on $f_{NL}$ (vi). In \fig{H10}, the bound for large $m$ is the gravitational decay width (vii) and in \fig{H12} it is the lightness of the curvaton (ii). The low $m$ bound is the most interesting and is due to $V_T$ and $\Delta V$. Below this lower bound, these terms dominate the potential initially. However, in this region it is not possible to generate a sufficient amplitude of perturbations while satisfying all other constraints. If the terms $V_T$ and $\Delta V$ were ignored, the allowed region would fill the area outlined by the dotted lines. Thus, using the simplest, complete curvaton potential gives a substantially different parameter space than the assumption of a pure quadratic potential.

We also show the region of the parameter space which has high probability (assuming a long enough period of inflation). This is below the solid line marked ``F-P''. This high probability region is greatly reduced from the parameter space allowed by the constraints (i) to (viii) above and favours low $m$ and low $\Gamma$.

The dynamics of the curvaton in the allowed region can also be different from the original model. Some of the low $m$ allowed region is initially dominated by $V_T$. In addition, $\Delta V$ may also play a role in the dynamics, although it is subdominant. This would affect the calculation of the non-Gaussianity parameters $f_{NL}$ and $g_{NL}$. In a later paper, we expect to produce a full numerical analysis of the parameter space, including predictions for $f_{NL}$ and $g_{NL}$. We expect to be able to carefully explore the whole curvaton parameter space, including all allowed values of $H_\ast$. We will also consider the corrections to other curvaton models.

In summary, we have completed the simplest curvaton model by including the full effect of the required decay of the curvaton. This gives rise to an additional term in the effective potential of the curvaton, proportional to $\sigma^4\log\sigma$. A thermal correction to the potential should also be considered. Using an analytical approximation, we were able to calculate the parameter space for this model consistent with various constraints, such as the observational limit on $f_{NL}$. We showed that the parameter space calculated for the original curvaton model $V_0$ is greatly reduced when the thermal correction and the correction $\Delta V$ due to the curvaton's coupling are included. Computations of curvaton parameters using only $V_0$ are not necessarily valid due to the differing dynamics of the curvaton model. They should therefore be repeated with the realistic potential presented in this letter.

A particularly large effect on the available parameter space comes from consideration of the initial conditions for the curvaton's evolution. Assuming a long enough period of inflation, we calculated the probability of the curvaton having the correct initial conditions in order to match the observed curvature perturbation. The high probability region is substantially reduced compared to the parameter space neglecting to consider the initial conditions. In conclusion, it is essential to consider the full potential for the curvaton as this can have a dramatic effect on the predictions of the model.

OT acknowledges the support by the Magnus Ehrnrooth foundation. RL and KE are respectively supported by the Academy of Finland grants 131454 and 218322. We wish to thank Tomo Takahashi for useful comments on an earlier version of this manuscript.

\end{document}